\documentclass[aps,preprint]{revtex4}

\usepackage{epsfig}
\usepackage[vflt]{floatflt}
\begin{document}

\title{Dynamics of the Bose-Einstein condensation:\\
analogy with the collapse dynamics of a classical self-gravitating
Brownian gas}

\author{Julien Sopik, Cl\'ement Sire and Pierre-Henri
Chavanis}

%\begin{center}
\affiliation{Laboratoire de Physique Th\'eorique (UMR 5152 du CNRS),
Universit\'e Paul Sabatier\\
118, route de Narbonne, 31062 Toulouse Cedex 4,
France\\
E-mail : {\it Sopik/Clement.Sire/Chavanis@irsamc.ups-tlse.fr}}
\vspace{0.5cm}
%\end{center}

%----------------------------------------

\begin{abstract}
We consider the dynamics of a gas of free bosons within a
semi-classical Fokker-Planck equation for which we give a physical
justification. In this context, we find a striking similarity
between the Bose-Einstein condensation in the canonical ensemble,
and the gravitational collapse of a gas of classical
self-gravitating Brownian particles. The paper is mainly devoted
to the complete study of the Bose-Einstein ``collapse'' within
this model. We find that at the Bose-Einstein condensation
temperature $T_c$, the chemical potential $\mu(t)$ vanishes
exponentially with a universal rate that we compute exactly. Below
$T_c$, we show analytically that $\sqrt{\mu(t)}$ vanishes linearly
in a finite time $t_{coll}$. After $t_{coll}$, the mass of the
condensate grows linearly with time and saturates exponentially to
its equilibrium value for large time. We also give analytical
results for the density scaling functions, for the corrections to
scaling, and for the exponential relaxation time. Finally, we find
that the equilibration time (above $T_c$) and the collapse time
$t_{coll}$ (below $T_c$), both behave like $-T_c^{-3}\ln|T-T_c|$,
near $T_c$.
\end{abstract}
\maketitle

\section{Introduction}

The Bose-Einstein condensation is a fundamental result of quantum
statistics \cite{bc}. Below a critical temperature $T_c$, a finite
fraction of bosons enters the ground state and a Bose-Einstein
condensate (BEC) forms \cite{books}. Observation of Bose-Einstein
condensation in cold atomic samples has been first reported in 1995 by
three different groups \cite{rb87,li26,na23} in a vapor of
spin-polarized $^{87}{\rm Rb}$, $^{7}{\rm Li}$ and $^{23}{\rm Na}$
atoms. Apart from laboratory experiments, another interesting
application of Bose condensation is related to the problem of boson
star formation from the dark matter in the Universe \cite{bstar}.  In
that case, Bose star formation involves the axion as a dark matter
particle candidate \cite{axion}.

Since its discovery \cite{bose,einstein}, several authors have
attempted to develop kinetic models to describe the dynamical
process of the Bose-Einstein condensation. The dynamical evolution
of a homogeneous gas of bosons interacting only via ``collisions''
can be studied using an appropriate form of the Boltzmann equation
\cite{nordheim,bloch,uu} which takes into account the specificities of
the Bose statistics. Since the gas is isolated and the energy
conserved, this model describes a {\it microcanonical} situation.  For
this model, the Bose-Einstein condensation in momentum space has been
considered by Semikoz \& Tkachev \cite{st} and Lacaze et {\it al.} 
\cite{lacaze}.  Alternatively, the {\it canonical} evolution of a
system of non interacting bosons coupled to a thermostat imposing the
temperature can be described by a semi-classical Fokker-Planck
equation
\cite{kompaneets,kan1,kan2}. For $T>T_{c}$, where $T_c$ is the
critical temperature, this equation converges towards the
Bose-Einstein distribution. The main purpose of this paper is a
complete study of this equation below the critical temperature
$T_{c}$, in order to understand the dynamics of the formation of the
condensate. Our study is restricted to a spatially homogeneous system
without interaction at arbitrary temperature.  Alternatively, the
collapsing dynamics of a trapped Bose-Einstein condensate (BEC) with
attractive interaction is often analyzed in terms of the
Gross-Pitaevskii (GP) or nonlinear Schr\"odinger equation (see, for
instance \cite{rasmussen}). This describes the formation of a
spatially inhomogeneous condensate at $T=0$. The GP equation can
display a self-similar collapse, but it occurs in position space while
the system that we shall consider is spatially homogeneous and the
condensation occurs in ${\bf k}$-space.

We thus consider a population of free bosons in $d$ dimensions with
dispersion relation $\varepsilon({\bf k})=\frac{k^2}{2m}$ (we set
$m=1$ in the following). This system is assumed to be strongly coupled
to a thermal bath at temperature $T$. Initially, at $t=0$, the system
is prepared in an initial state which can be, for instance, the
equilibrium distribution at some high temperature $T_0\gg T$.  At
infinite time, we expect the system to reach thermal equilibrium at
temperature $T$, characterized by the Bose-Einstein distribution if $T>T_c$
(where $T_c$ is the Bose-Einstein condensation temperature)
\begin{equation}
\rho_{BE}({\bf k})=\frac{1}{\exp\left(\frac{\beta
k^2}{2}+\mu\right)-1}, \label{rhoeq1}
\end{equation}
where $\mu$ is the chemical potential. Below $T_c$, the equilibrium
distribution is the Bose-Einstein distribution with $\mu=0$ plus a
Dirac peak at ${\bf k}={\bf 0}$ (the condensate) containing the rest
of the mass.

\subsection{Bosonic Fokker-Planck Equation}
In this paper, we are interested in the temporal evolution of the
occupation number at momentum ${\bf k}$, that we call $\rho({\bf
k},t)$, from an arbitrary initial state to the final equilibrium
state described above. If our particles were classical instead of
bosons (of course no condensate can appear in that case), the
coupling to the thermal bath can be modelized by a random force
and a friction, and the evolution equation for the momentum of a
given particle is described by the Langevin dynamics
\begin{equation}
\label{lang}
\frac{d{\bf k}}{d t}=-\frac{{\bf k}}{\xi}+{\bf f}(t),
\end{equation}
where ${\bf f}(t)$ is a delta-correlated random force, whose
components satisfy
\begin{equation}
\langle{f}_i(t){f}_j(t')\rangle=2 D\delta_{ij}\delta(t-t'),\quad
i,j\in[1,...\,,d].
\end{equation}
In order to recover the equilibrium equipartition theorem, we must
impose the Einstein relation
\begin{equation}
D=\frac{T}{\xi},
\end{equation}
where we have set the Boltzmann constant $k_{B}=1$. The Fokker-Planck
equation describing the temporal evolution of the momentum distribution reads
\begin{equation}
\frac{\partial\rho}{\partial t}=\frac{1}{\xi}\nabla_{\bf k}\biggl \lbrack
T\nabla_{\bf k}\rho+\rho {\bf k}\biggr\rbrack. \label{lange}
\end{equation}
Ultimately, the momentum distribution converges to the expected
Boltzmann distribution
\begin{equation}
\rho_{B}({\bf k})=Z^{-1}\exp\left(-\frac{\beta
k^2}{2}\right).
\end{equation}
Now, coming back to the Bose gas, we ask whether a stochastic dynamics
can be introduced, which accurately describes the actual
evolution \footnote{In a classical Monte-Carlo simulation using local updating (for
instance in an Ising spin system, $s_i\to -s_i$ with Glauber or
Metropolis probabilities), the Monte-Carlo time, as measured as the
number of updates, has in fact a physical relevance. In simulations of soap bubbles using the Potts model (a
generalization of the Ising model) \cite{soapnum}, the Monte-Carlo
time is found to be simply proportional to the actual time, and these
times can be made equal by a proper choice of $\xi$. Numerically
\cite{soapnum}, theoretically \cite{soapth1,soapth2,soapth3} and in
real physical systems \cite{soapexp}, the typical area of a bubble is
found to grow linearly with time. The out of
equilibrium dynamics of the Ising model quenched from high temperature
to a temperature below the ferromagnetic critical temperature, leads
to a coarsening dynamics where domains of $\uparrow$ and $\downarrow$
spins grow on a scale $L(t)$ \cite{isingth1,isingth2}.  An experiment
in nematic liquid crystals exactly mimics this situation
\cite{isingexp}, where the physical time is found to be indeed
proportional to the Monte-Carlo time in simulations.  One obtains
$L(t)\sim t^{1/2}$, and spin temporal correlations are found to have
the same functional form \cite{isingth1,isingth2,isingexp}. It is
clearly comforting and satisfying that the theoretical dynamics
modelizing the coupling to a thermal bath leading to equilibrium
actually describes the evolution of out of equilibrium physical
systems. However in {\it quantum systems}, it is not clear at all how
to relate the simulation time (number of world lines affected by the
Monte-Carlo dynamics) to the actual time.}. Restraining ourselves to
non interacting bosons, we shall see that one can give a reasonable
answer to this question.  This will permit the description of the
dynamical formation of the condensate at and below $T_c$, which is the
main purpose of this paper.

Several authors have considered this problem \cite{kan1,kan2}. Let
us introduce a stochastic dynamics, by defining the rate at which
particles with momentum ${\bf k}$ get a new momentum ${\bf k'}$.
Following Kaniadakis {\it et al.} \cite{kan1}, we assume the
following form of this rate
\begin{equation}
W({\bf k}\to {\bf k'})=w({\bf k},{\bf k-k'})\,a[\rho({\bf k},t)]\,
b[\rho({\bf k'},t)].
\end{equation}
Contrary to classical dynamics, we assume a dependence of $W({\bf
k}\to {\bf k'})$ with $\rho({\bf k},t)$ and $\rho({\bf k'},t)$.
Otherwise, we would simply recover the Fokker-Planck equation
Eq.~(\ref{lange}), as will be shown below. In fact, $W({\bf k}\to
{\bf k'})$ being a {\it rate} of departure from the state of
momentum ${\bf k}$, it seems reasonable to assume that it  should
not depend on the population of this initial state. Hence, we set
$a(\rho)=1$. Then, one can write the master equation describing
the time evolution of the particle distribution as
\begin{equation}
\frac{\partial\rho}{\partial t}=\int\left[\rho({\bf k'},t)W({\bf
k'}\to {\bf k})-\rho({\bf k},t)W({\bf k}\to {\bf k'})\right]
d{\bf k'}.\label{master}
\end{equation}
Assuming that the evolution is sufficiently slow, and local, such
that the dynamics only permits values of ${\bf k'}$ close to ${\bf
k}$, one can develop $W({\bf k}\to {\bf k'})$ in powers of ${\bf
k-k'}$ in Eq.~(\ref{master}). After some algebra, and proceeding
along the line of \cite{kan1}, one obtains a Fokker-Planck-like
equation
\begin{equation}
\frac{\partial\rho}{\partial t}=\nabla_{\bf k}\biggl \lbrack
b(\rho)\nabla_{\bf k}(D\rho)+\rho b(\rho){\bf J}-D\rho
b'(\rho)\nabla_{\bf k}\rho\biggr\rbrack, \label{fkbose1}
\end{equation}
where assuming isotropy, we obtain
\begin{eqnarray}
\label{br1}
D({\bf k})&=&\frac{1}{2d}\int w({\bf k},\Delta {\bf k})(\Delta{\bf
k})^2\,d(\Delta {\bf k}),\\
{\bf J}({\bf k})&=&-\int w({\bf k},\Delta {\bf k})\Delta {\bf k}\,d(\Delta {\bf k}).
\end{eqnarray}
We now make the simplification that the diffusion coefficient $D({\bf
k})$ is momen\-tum-in\-de\-pen\-dent, and that the current ${\bf J}$
is simply proportional to the particles velocity or momentum ${\bf
k}$:
\begin{eqnarray}
\label{br2}
D({\bf k})&=&\frac{T}{\xi},\\
{\bf J}({\bf k})&=&\frac{{\bf k}}{\xi}.\label{br2b}
\end{eqnarray}
Going from Eq.~(\ref{br1}) to Eq.~(\ref{br2}) is a standard assumption in
kinetic theory. It is exactly achieved when $w({\bf k},
\Delta{\bf k})$ is a symmetric analytic function of $(\Delta{\bf
k}+{\bf k}\Delta t/\xi)/\sqrt{\Delta t}$, where $\Delta t$ is the
discretized time step (ultimately $\Delta t\to 0$). This function must
also decay fast enough so that its two first moments are well
defined. In the case of the usual Brownian motion based on the
Langevin equations (\ref{lang}), this function is simply a
Gaussian. Using Eqs.~(\ref{br2}) and (\ref{br2b}), Eq.~(\ref{fkbose1}) becomes
\begin{equation}
\frac{\partial\rho}{\partial t}=\frac{1}{\xi}\nabla_{\bf k}\biggl
\lbrack T(b(\rho)-\rho b'(\rho))\nabla_{\bf k}\rho+\rho
b(\rho){\bf k}\biggr\rbrack. \label{fkbose2}
\end{equation}
Note that choosing $b(\rho)=1$, {\it i.e.} a transition rate which
does not depends on the population of the new momentum state,
leads to the standard Fokker-Planck equation Eq.~(\ref{lange}) for
classical particles. In the general case, the stationary
distribution for ${\bf k}\ne {\bf 0}$ satisfies
\begin{equation}
\left(\frac{1}{\rho}- \frac{b'(\rho)}{b(\rho)}\right) \nabla_{\bf
k}\rho=-\beta {\bf k},
\end{equation}
which can be integrated, leading to
\begin{equation}
\frac{\rho}{b(\rho)}=\exp\left(-\frac{\beta k^2}{2}-\mu\right),
\end{equation}
where $\mu$ is an integration constant. In order to recover the
Bose-Einstein distribution Eq.~(\ref{rhoeq1}), we must make the
unique choice
\begin{equation}
b(\rho)=1+\rho.
\end{equation}
We conclude that the classical rate is amplified by the factor
$b(\rho)>1$, which translates the tendency of bosons to fill
already occupied states. Note that the choice $b(\rho)=1-\rho$
leads to the Fermi-Dirac distribution \cite{kan2}, where $b(\rho)$
now suppresses the probability to hop to an already occupied
state, as can be expected for fermions (note that a generalized
Fokker-Planck equation incorporating an exclusion principle has
been introduced independently in \cite{csr} in the very different
context of the violent relaxation of collisionless stellar systems
and 2D vortices). This approach has been extended to the case of
intermediate statistics, associated to $b(\rho)=1+\kappa\rho$
($\kappa\in[-1,1]$) \cite{kan2}. It can be also generalized in
order to recover Tsallis statistical thermodynamics at equilibrium
from a general kinetic equation \cite{kan2,cstsallis}.

Focusing on the bosonic case, we finally obtain the associated
Fokker-Planck equation
\begin{equation}
\frac{\partial\rho}{\partial t}=\frac{1}{\xi}\nabla_{\bf k}\biggl
\lbrack T\nabla_{\bf k}\rho+\rho(1+\rho){\bf k}\biggr\rbrack.
\label{bose1}
\end{equation}
This bosonic Kramers equation could be directly obtained
from a modified Langevin equation
\begin{equation}
\frac{d{\bf k}}{d t}=-\frac{{\bf k}}{\xi}(1+\rho({\bf k},t))+{\bf
f}(t),
\end{equation}
where the friction term which tends to move the momentum toward
${\bf k}={\bf 0}$ now increases with the occupation number at
momentum ${\bf k}$. In this context, the origin of the
Bose-Einstein instability is quite clear.

Finally, for later reference, we write the bosonic Fokker-Planck
equation for a spherically symmetric solution. Setting
$\xi=1$, we get:
\begin{equation}
\frac{\partial\rho}{\partial t}=T \left(\frac{\partial^2
\rho}{\partial k^2} +\frac{d-1}{k}\frac{\partial \rho}{\partial
k}\right) +d\rho(1+\rho)+k(2\rho+1)\frac{\partial \rho}{\partial
k}. \label{bose2}
\end{equation}
For the integrated density
\begin{equation}
M(k,t)=\int_{0}^{k}\rho(k',t)k'^{d-1}\,dk', \label{normgrav}
\end{equation}
we obtain
\begin{equation}
\frac{\partial M}{\partial t}=T \left(\frac{\partial^2 M}{\partial
k^2} -\frac{d-1}{k}\frac{\partial M}{\partial k}\right)
+k\frac{\partial M}{\partial
k}\left(\frac{1}{k^{d-1}}\frac{\partial M}{\partial k}+1\right).
\label{bose3}
\end{equation}

It has to be noted that the bosonic Fokker-Planck equation
(\ref{bose1}) valid for non-interacting bosons in contact with a heat
bath (canonical ensenble) is the counterpart of the bosonic Boltzmann
equation \cite{nordheim,bloch,uu} valid for an isolated system of
interacting bosons (microcanonical ensemble). These kinetic equations
are {\it semi-classical} equations where quantum mechanics enters only
through the Bose-Einstein statistics. The bosonic Boltzmann equation
can be derived from a fully quantum treatment as discussed in
Sec. \ref{sec_comp}. The justification of the bosonic Fokker-Planck
equation from a fully quantum mechanics treatment would be interesting but is
beyond the scope of this paper.

\subsection{Analogy with a self-gravitating gas of Brownian
particles}
\label{sec_brown}

In a series of recent papers
\cite{cstsallis,charosi,csdim,cspostcoll,cstcoll}, two of the
present authors have introduced and systematically studied the
dynamical properties of a self-gravitating gas of Brownian particles
in all spatial dimensions. This is the {\it canonical} version of the
original and certainly more challenging problem of self-gravitating
Newtonian particles in the {\it microcanonical} ensemble. In the
overdamped limit, the Langevin equation for the position in real space
of a particle reads
\begin{equation}
\frac{d{\bf r}}{d t}=-\frac{\nabla\Phi}{\xi}+{\bf f}(t),
\end{equation}
where the gravitational potential $\Phi({\bf r},t)$ must be
computed self-consistently using the Poisson equation
\begin{equation}
\Delta\Phi({\bf r},t)=S_{d}G\rho({\bf r},t),
\end{equation}
where $S_d$ is the surface of the unit $d$-dimensional sphere, and
$G$ is Newton's constant. The associated Fokker-Planck-Poisson
(or Smoluchowski-Poisson) system is obtained straightforwardly
\begin{eqnarray}
\frac{\partial\rho}{\partial t}&=&\frac{1}{\xi}\nabla\biggl
\lbrack T\nabla\rho+\rho\nabla\Phi\biggr\rbrack,
\label{grav1}\\
\Delta\Phi&=&S_{d}G\rho.\label{grav1b}
\end{eqnarray}
As discussed in \cite{ribot}, these equations also describe the chemotaxis of
bacterial populations in biology, by a proper re-interpretation of the
parameters. From now on, we get rid of non essential constants by
setting
\begin{equation}
G=\xi=S_d=1.
\end{equation}
For a time-dependent radial solution, the system of equations
Eqs.~(\ref{grav1},\ref{grav1b}) can be put into a unique equation
\cite{csdim}
\begin{equation}
\frac{\partial\rho}{\partial t}=T \left(\frac{\partial^2
\rho}{\partial r^2} +\frac{d-1}{r}\frac{\partial \rho}{\partial
r}\right) +\rho^2+\frac{M}{r^{d-1}}\frac{\partial \rho}{\partial
r}, \label{grav2}
\end{equation}
where
\begin{equation}
M(r,t)=\int_{0}^{r}\rho(r')r'^{d-1}\,dr', \label{normgravb}
\end{equation}
is the integrated density. Actually, the equation for $M(r,t)$ looks
even simpler
\begin{equation}
\frac{\partial M}{\partial t}=T \left(\frac{\partial^2 M}{\partial
r^2} -\frac{d-1}{r}\frac{\partial M}{\partial r}\right)
+\frac{M}{r^{d-1}}\frac{\partial M}{\partial r}. \label{grav3}
\end{equation}
It is clear that Eqs.~(\ref{grav2},\ref{grav3}) are strikingly
similar to the dynamical equations found in the context of the
Bose-Einstein condensation Eqs.~(\ref{bose2},\ref{bose3}). Apart
from the identical diffusion term, we note that the non linear
terms have the same dimension as $\rho^2$ in
Eqs.~(\ref{bose2},\ref{grav2}) and dimension $M^2{\times} (k,r)^{-d}$ in
Eqs.~(\ref{bose3},\ref{grav3}) (we shall see later that the $+1$
term at the end of Eq.~(\ref{bose3}) is essentially irrelevant as
far as the dynamics of the Bose-Einstein condensation at or below
$T_c$ is concerned).

Before summarizing our results concerning the dynamics of a gas of
bosons in $d=3$ described by Eqs.~(\ref{bose2},\ref{bose3}), we
would like to mention only the main results obtained for a
self-gravitating Brownian gas obeying
Eqs.~(\ref{grav2},\ref{grav3}). The comparison with this
surprisingly close model will certainly prove interesting.
\begin{itemize}
\item In $d\geq 3$, Eqs.~(\ref{grav2},\ref{grav3}) reproduce the
known equilibrium profile for $T\geq T_c$. For $T<T_c$, a
dynamical instability arises coinciding with the absence of
minimum of the free energy (thermodynamical instability)
\cite{charosi,csdim}.

\item In $d\geq 3$, for $T<T_c$, the density develops a scaling
profile $\rho(r,t)=\rho_0 f(r/r_0)$, with $\rho_0(t)=Tr_0^{-2}(t)\sim
(t_{coll}-t)^{-1}$ and $f(x)\sim x^{-2}$ for $x\rightarrow
+\infty$. Hence, the central density diverges in a finite time
$t_{coll}$. Note that $f$ can be calculated analytically in all
dimensions, and that corrections to scaling have been evaluated
\cite{charosi,csdim}.

\item In $d\geq 3$, for $T<T_c$, and for $t\geq t_{coll}$, the
central density is swallowed by an emerging condensate at ${\bf
r}={\bf 0}$, which grows like $M_0(t)\sim (t-t_{coll})^{d/2-1}$.  In
this post-collapse regime, the residual density obeys a backward
dynamical scaling, $\rho(r,t)=\rho_0 g(r/r_0)$, with
$\rho_0(t)=Tr_0^{-2}(t)\sim (t-t_{coll})^{-1}$ and $g(x)\sim x^{-2}$
for $x\rightarrow +\infty$. For large time, the central condensate
saturates exponentially to the {\it total} mass, whereas the residual
density vanishes exponentially with the same rate, which has been
calculated by semi-classical technics
\cite{cspostcoll}. We found that the relaxation time $\tau$ above $T_c$
diverges like $\tau\sim K_+(T-T_c)^{-1/2}$, where $K_+$ is known
exactly. Approaching $T_c$ from below, the collapse time
$t_{coll}$ diverges like $t_{coll}\sim K_-(T_c-T)^{-1/2}$, where
again $K_-$ has been computed analytically \cite{cstcoll}.

\item In $d=2$, at $T_c$, the density obeys a dynamical scaling
with a known scaling function $f$. The central density diverges like
$\rho_0(t)\sim c_1\exp(c_2\sqrt{t})$, where $c_1$ and $c_2$ are known
universal constants \cite{csdim,banb} (in an unbounded domain, the
divergence is logarithmic instead of exponential \cite{virial}). Below
$T_c$, the collapse dynamics occurs after a finite time $t_{coll}$. At
$t_{coll}$, a fraction $T/T_c$ of the total mass has condensed at
${\bf r}={\bf 0}$. Using the Virial theorem, we show that all the mass
must collapse at ${\bf r}={\bf 0}$ in the post-collapse regime
\cite{virial}. We mention these results in $d=2$, because the dynamics
of the Bose-Einstein condensation in $d=3$ will share some qualitative
analogy with this case.
\end{itemize}

We thus expect a surprising parallel between the collapse dynamics of
a classical self-gravitating Brownian gas, which occurs when the
kinetic pressure $P\sim T$ is not strong enough to counterbalance the
gravitational attraction, and a bosonic gas, which forms a condensate
at ${\bf k}={\bf 0}$, when the usual Bose-Einstein distribution is not
able to accommodate for all the mass, even at zero chemical potential
(see Section II.A). However, we note one fundamental difference
between these two systems.  For $t\rightarrow +\infty$, strictly below
$T_c$ all the mass collapses at ${\bf r}={\bf 0}$ in the gravitational
case, whereas the mass of the condensate is temperature-dependent in
the case of the Bose-Einstein transition.

\subsection{Summary of results}

In Section II.A and II.B, we briefly review some basic results
concerning the static properties of the Bose-Einstein
condensation, and introduce a simpler model (SM), for which the
static and dynamical properties can be analytically studied, and
which will be equivalent in the dense region to the original
Bose-Einstein model (BEM) described by
Eqs.~(\ref{bose2},\ref{bose3}). In section II.C, we show that
Eqs.~(\ref{bose2},\ref{bose3}) maximize the rate of dissipation of
bosonic free energy, and that the dynamical instability exactly
coincides with the thermodynamical instability giving rise to the
Bose-Einstein condensation below $T_c$.

In Section III, we address the collapse dynamics at and below
$T_c$. In Section III.A, we compute the density scaling profile
which is found to be independent of the temperature $T\leq T_c$,
up to a temperature-dependent multiplicative factor. In Section
III.B, we show that at $T=T_c$ the central density (at ${\bf
k}={\bf 0}$) diverges exponentially with time, with a rate which
can be calculated analytically. We compute the exponentially
decaying corrections to the final Bose-Einstein distribution with
$\mu=0$. In Section III.C, we show that for $T<T_c$, the chemical
potential vanishes in a finite time and that $\sqrt{2T\mu(t)}\sim
c(T)(t_{coll}-t)$, where $c(T)$ is computed analytically close to
$T_c$. We show that $t_{coll}$ diverges logarithmically as $T$
approaches $T_c$ from below. In Section III.D, we address the
specific collapse dynamics obtained strictly at $T=0$.

In Section IV, we study the post-collapse dynamics arising for $T<
T_c$, and for times $t>t_{coll}$. We show that the condensate mass
initially grows linearly, with a slope which can be calculated
exactly near $T_c$. For large times, the mass of the condensate
saturates exponentially fast to its equilibrium value, with a rate
analytically known close to $T_c$.

In Section V, we consider the relaxation time above $T_c$, which
is computed in different limits. This relaxation time does not
diverge at $T_c$, but the time after which this relaxation regime
occurs (that we call the equilibration time) diverges
logarithmically as the temperature approaches $T_c$ from above.

\section{Equilibrium properties and a simpler model in a box}

\subsection{Bose-Einstein condensation for a gas of free bosons}

Let us briefly repeat the few steps leading to the Bose-Einstein
condensation. Considering free bosons, the equilibrium
occupation number at momentum ${\bf k}$ is
\begin{equation}
\rho_{BE}({\bf k})=\frac{1}{\exp\left(\frac{\beta
k^2}{2}+\mu\right)-1}. \label{rhoeq}
\end{equation}
The total mass is fixed to unity, and the chemical potential $\mu$
is defined implicitly by the total mass constraint (for
simplicity, we set the geometrical factor $S_d=1$, which amounts
to fixing $M=S_d=1$)

\begin{equation}
M=\int_{0}^{\infty}\rho_{BE}(k)k^{d-1}\,dk=1, \label{normbose}
\end{equation}
which can be rewritten as the identity
\begin{equation}
\beta^{d/2}=\int_{0}^{\infty}\frac{k^{d-1}}{\exp\left(\frac{
k^2}{2}+\mu\right)-1}\,dk.
\end{equation}
Below a certain temperature $T_c$, this distribution cannot include
all the mass. The critical temperature corresponds to
$\mu=0$. Specializing to the case of dimension $d=3$ which is the
focus of this paper (there is no condensation for $d<3$), we find
\begin{equation}
\beta_c=\left(\frac{\pi}{2}\right)^{1/3}\zeta^{2/3}
\left(\frac{3}{2}\right),\quad \beta_c\approx 2.20494...
\end{equation}
where $\zeta(x)=\sum_{k=1}^{+\infty}k^{-x}$ is the Riemann
$\zeta$-function. Below $T_c$, the momentum distribution is the
Bose-Einstein distribution with $\mu=0$ plus a Dirac peak at ${\bf
k}={\bf 0}$ containing the excess of mass $M_0$, with
\begin{equation}
M_0=1-\left(\frac{\beta_c}{\beta}\right)^{3/2}
\sim\frac{3}{2}\frac{T_c-T}{T_c}, \quad\frac{T-T_c}{T_c}\ll 1.
\end{equation}
Above $T_c$, we find the following asymptotics which will be
useful later:
\begin{equation}
\mu\sim\frac{9}{8\pi^2}T_c^{-3}
\left(\frac{T-T_c}{T_c}\right)^2,\quad \frac{T-T_c}{T_c}\ll 1,
\label{mu1}
\end{equation}
and
\begin{equation}
\mu\sim\frac{3}{2}\ln T, \quad T\to +\infty.
\end{equation}

\subsection{A simpler model in a box}

We now introduce a simpler model (hereafter denoted SM), which
will share the same properties as our original model, as far as
the condensation dynamics properties are concerned. In the kinetic
equation Eq.~(\ref{bose2}), we replace $(\rho+1)$ with $\rho$,
which should be valid in the dense region around ${\bf k}={\bf
0}$, during the formation of the condensate. We obtain
\begin{equation}
\frac{\partial\rho}{\partial t}=\nabla_{\bf k}\cdot \biggl \lbrack
T\nabla_{\bf k}\rho+\rho^2 {\bf k}\biggr\rbrack, \label{mod1}
\end{equation}
and the following equation for a radial solution
\begin{equation}
\frac{\partial\rho}{\partial t}=T \left(\frac{\partial^2
\rho}{\partial k^2} +\frac{d-1}{k}\frac{\partial \rho}{\partial
k}\right) +d\rho^2+2k\rho\frac{\partial \rho}{\partial k}.
\label{mod2}
\end{equation}
The integrated density satisfies
\begin{equation}
\frac{\partial M}{\partial t}=T \left(\frac{\partial^2 M}{\partial
k^2} -\frac{d-1}{k}\frac{\partial M}{\partial k}\right)
+\frac{1}{k^{d-2}}\left(\frac{\partial M}{\partial k}\right)^2,
\label{mod3}
\end{equation}
which is even more similar to the dynamical equation
Eq.~(\ref{grav3}) for a gas of self-gravitating Brownian particles
than Eq.~(\ref{bose3}). The stationary solution can be easily
calculated
\begin{equation}
\rho({\bf k})=\frac{1}{\frac{\beta k^2}{2}+\mu}, \label{rhoeqmod}
\end{equation}
which amounts to replacing the function $\exp x$ is the
Bose-Einstein distribution by the function $(1+x)$. Note that the
distribution (\ref{rhoeqmod}) is the Lorentzian. Since this
distribution $\rho({\bf k})$ is not integrable up to infinity in
$d\geq 2$, we introduce a momentum cut-off
\begin{equation}
k\leq \Lambda.
\end{equation}
As before, the total mass constraint reads
\begin{equation}
M=\int_{0}^{ \Lambda}\frac{k^{d-1}}{\frac{\beta
k^2}{2}+\mu}\,dk=1.
\end{equation}
From now on, we set $d=3$, as well as in the rest of the paper. Then,
the integrated  density can be explicitly calculated
\begin{equation}
M(k)=2T\left(k-\sqrt{2\mu T}\arctan\left(\frac{k}{\sqrt{2\mu
T}}\right)\right).
\end{equation}
The density profile Eq.~(\ref{rhoeqmod}) can accommodate for all
the mass up to the temperature
\begin{equation}
\beta_c=2\Lambda.
\end{equation}
Below $T_c$, a Dirac peak appears at ${\bf k}={\bf 0}$, containing
the excess mass
\begin{equation}
M_0=\frac{T_{c}-T}{T_c}.
\end{equation}
Just above $T_c$, the chemical potential vanishes in a manner
similar to the original Bose-Einstein model (BEM),
\begin{equation}
\mu\sim\frac{1}{2\pi^2}T_c^{-3}
\left(\frac{T-T_c}{T_c}\right)^2,\quad \frac{T-T_c}{T_c}\ll 1,
\label{mu2}
\end{equation}
whereas at high temperatures, the two models are qualitatively
different, as the particles momentum cannot spread up to infinity
in the SM. Hence, the chemical potential converges at high
temperatures
\begin{equation}
\mu\to\frac{\Lambda^3}{3}, \quad T\to +\infty.
\end{equation}

\subsection{Some properties of the bosonic Fokker-Planck equation}

Equation (\ref{bose1}) belongs to a general class of nonlinear
Fokker-Planck (NFP) equations considered by Kaniadakis
\cite{kan2}, Frank \cite{frank} and Chavanis \cite{gen}. These
nonlinear Fokker-Planck equations arise when the coefficients of
diffusion, mobility and friction depend explicitly on the
distribution function. There are different ways to write these NFP
equations. One convenient form for our present purposes is
\cite{gen}:
\begin{equation}
\frac{\partial\rho}{\partial t}=-\nabla\cdot {\bf J}_{*}= \nabla\cdot
\left \lbrack \frac{1}{\xi}\left
(T\nabla\rho+\frac{1}{C''(\rho)}\nabla\Phi\right )\right \rbrack,
\label{nfpconvenient}
\end{equation}
where $C(\rho)$ is a convex function and $\xi>0$ can depend on ${\bf
r}$ and $t$, e.g.  on $\rho({\bf r},t)$. Here, $\Phi({\bf r})$ is a
fixed external potential but we can also consider the situation where
$\Phi({\bf r},t)$ is generated by $\rho({\bf r},t)$ \cite{gen}, like
in the case of self-gravitating Brownian particles (see Section
\ref{sec_brown}). Taking
$C(\rho)=\rho\ln\rho-\frac{1}{\kappa}(1+\kappa \rho)\ln
(1+\kappa\rho)$, we recover the bosonic Fokker-Planck equation
(\ref{bose1}) for $\kappa=+1$, the fermionic Fokker-Planck equation
for $\kappa=-1$ and the classical Fokker-Planck equation for
$\kappa=0$ (for other values of $\kappa$, this describes intermediate
``quon'' statistics). For the simpler model Eq.~(\ref{mod1}), we have
$C(\rho)=-\ln\rho$ describing logotropes \cite{log}. Introducing a
generalized free energy
\begin{equation}
F=E-TS=\int \rho\Phi \,d{\bf r}+T\int C(\rho)\,d{\bf r},
\label{gfree}
\end{equation}
one can show that $F(t)$ is monotonically decreasing: $\dot
F=-\int \xi C''(\rho) {\bf J}_{*}^{2}\, d{\bf r}\le 0$. Therefore,
if $F$ is bounded from below, the density $\rho({\bf r},t)$ will
converge, for $t\rightarrow +\infty$, to the stationary solution
$\rho_{eq}({\bf r})$ such that ${\bf J}_{*}={\bf 0}$ (in that case
$F$ is called a Lyapunov functional). The stationary solution of
Eq.~(\ref{nfpconvenient}) is determined by
\begin{equation}
\rho_{eq}({\bf r})=(C')^{-1}\lbrack -\beta\Phi({\bf r})-\alpha \rbrack,
\label{rhoeqc}
\end{equation}
where $\beta=1/T$ and $\alpha$ is an integration constant. This
stationary solution is a critical point of free energy
Eq.~(\ref{gfree}) at fixed mass, {\it i.e.} it satisfies $\delta
F+\alpha T\delta M=0$, where $\alpha T$ is a Lagrange multiplier.
Moreover, it is shown in \cite{gen} that a stationary solution of
Eq.~(\ref{nfpconvenient}) is linearly dynamically stable if, and
only if, it is a {\it minimum} of $F$ at fixed mass. In fact, when
$\Phi({\bf r})$ is an external potential, the critical points of
$F$ at fixed mass are necessarily minima since
$\delta^{2}F=\frac{1}{2}T\int C''(\rho)(\delta\rho)^{2}\,d{\bf r}\ge
0$. Therefore, if a critical point of free energy exists it is the
only minimum of $F$ and, consequently, $F$ is bounded from below:
$F[\rho]\ge F[\rho_{eq}]$. In that case, the dynamical equation
Eq.~(\ref{nfpconvenient}) will relax towards $\rho_{eq}({\bf r})$
for $t\rightarrow +\infty$.

The NFP equation Eq.~(\ref{nfpconvenient}) can be justified in several
different ways \cite{gen}. It can be obtained from the linear
thermodynamics of Onsager by relating the current to the gradient of a
``chemical potential'' $\alpha({\bf r},t)=-\beta\Phi({\bf
r})-C'(\rho)$ that is uniform at equilibrium (see Eq.~(\ref{rhoeqc})),
writing ${\bf J}_{*}=1/\lbrack \beta \xi C''(\rho)\rbrack
\nabla\alpha$. The current can also be expressed as the functional
derivative of the free energy, writing ${\bf J}_{*}=-1/\lbrack \xi
C''(\rho)\rbrack\nabla (\delta F/\delta \rho)$ . Alternatively, the
NFP equation can be obtained by writing $\partial_{t}\rho=-\nabla\cdot
{\bf J}$ and looking for the optimal current ${\bf J}_{*}$ which
maximizes the rate of free energy dissipation $\dot F[{\bf J}]=\int
(TC''(\rho)\nabla\rho+\nabla\Phi){\bf J}\,d{\bf r}$ for a bounded
function $E_{d}[{\bf J}]=\frac{1}{2}\int \xi C''(\rho){\bf
J}^{2}\,d{\bf r}$ preventing arbitrarily large values of the current
${\bf J}$. This is the variational version of the linear
thermodynamics of Onsager. The optimal current, that is solution of
$\delta \dot F+\delta E_{d}=0$, is the one appearing in
Eq.~(\ref{nfpconvenient}) and it satisfies $\dot F[{\bf J}_{*}]=-2
E_{d}[{\bf J}_{*}]\le 0$. It indeed leads to the most negative value
of $\dot F$ (under constraints) since $\delta^{2}(\dot
F+E_{d})=\delta^{2}E_{d}=\frac{1}{2}\int \xi C''(\rho)(\delta {\bf
J})^{2}\,d{\bf r}\ge 0$.  These methods emphasize the (generalized)
thermodynamical structure of the NFP equation.

\section{Collapse dynamics}

\subsection{General scaling solutions for $T\leq T_c$}
\label{sec_gen}

We now consider the collapse dynamics occurring when the system is
suddenly quenched to $T_c$ or below $T_c$ from high temperature.
It is clear that both considered models should behave in a similar
manner near the dense region at ${\bf k}={\bf 0}$. Let us first
consider the condensation dynamics of the SM. We write
\begin{equation}
M(k,t)=2T\left(k-k_0\arctan\left(\frac{k}{k_0}\right)\right)
+F(k,k_0), \label{ansatz}
\end{equation}
where $k_0(t)$ is uniquely defined by the condition
\begin{equation}
\rho(k=0,t)=\frac{2T}{k_0^2},\label{rhok}
\end{equation}
ensuring that
\begin{equation}
F(k\to 0,k_0)\sim k^5.
\end{equation}
The normalization condition reads
\begin{equation}
M(k=\Lambda,t)=1=2T\left(\Lambda-k_0
\arctan\left(\frac{\Lambda}{k_0}\right)\right)+F(\Lambda,k_0).
\end{equation}
We now look for a scaling solution of the form
\begin{equation}
F(k,k_0)=k_0^\alpha f\left(\frac{k}{k_0}\right),
\end{equation}
and introduce the scaling variable
\begin{equation}
x=\frac{k}{k_0}.
\end{equation}
This leads to
\begin{equation}
M'(k,t)=k^2\rho(k,t)=\frac{2Tk^2}{k^2+k_0^2}+k_0^{\alpha-1}
f'\left(\frac{k}{k_0}\right),\label{Mprime}
\end{equation}
where, from now on, $M'(k,t)$ will denote the derivative with
respect to momentum. In analogy with the equilibrium density
Eq.~(\ref{rhoeqmod}), we observe that $\mu(t)=\beta k_0^2(t)/2$
can be seen as an effective chemical potential which is expected
to vanish as time increases.

In $d=3$, the dynamical equation for $M(k,t)$ is
\begin{equation}
\frac{\partial M}{\partial t}=T \left(M'' -\frac{2}{k}M'\right)
+\frac{M'^2}{k}. \label{mod3d3}
\end{equation}
Inserting the scaling ansatz, we find
\begin{eqnarray}
\frac{\partial M}{\partial t}&=&k_0^{\alpha-2}T
\left(f''+2\frac{x^2-1}{x(x^2+1)}f'\right)
+k_0^{2\alpha-3}\frac{f'^2}{x},\label{scaf}\\ &=&-2\dot
k_0T\left(\arctan(x)-\frac{x}{x^2+1}\right)+\dot k_0
k_0^{\alpha-1}\left(\alpha f-x f'\right).\nonumber
\end{eqnarray}
It is clear that for $\alpha=1$, all the terms of this equation scale
the same way, if one chooses $\dot k_0\sim k_0^{-1}$, i.e.
$k_{0}(t)\sim (t_{coll}-t)^{1/2}$.  However, we were able to prove
analytically that all the solutions of the resulting scaling equation
are non physical, leading to a density going to a constant for $k\gg
k_0$. In addition, when we wrote Eq.~(\ref{ansatz}), we implicitly
assumed that the second term of the right-hand side (RHS) is a
correction to the first one, which implies $\alpha >1$. Hence, we now
consider $\alpha > 1$, and matching the leading terms of
Eq.~(\ref{scaf}), we conclude that
\begin{equation}
\dot k_0=-c k_0^{\alpha-2},\label{dk0dt}
\end{equation}
and that the scaling function introduced above satisfies
\begin{equation}
f''+2\frac{x^2-1}{x(x^2+1)}f'=2c\left(\arctan(x)-\frac{x}{x^2+1}\right),
\end{equation}
which can be integrated once, leading to
\begin{equation}
f'(x)=\frac{2c}{3}\frac{x}{(x^2+1)^2}\left[
\left(x^4+6x^2-3\right)\arctan(x)+3x-2x^3-4x\ln\left(x^2+1\right)\right].
\label{fprime}
\end{equation}
Note that $f'(x)/x^2$ is exactly the scaling function for the
density as can be seen from Eq.~(\ref{Mprime}). We give below some
asymptotic results for $f(x)$ which will prove useful later:
\begin{eqnarray}
f(x)&=&\frac{\pi c}{6}x^2-\frac{4 c}{3}x+2\pi c\ln x+{\cal O}(1),
\quad x\to +\infty,\\
&=&\frac{2c}{15}x^5-\frac{8c}{45}x^7+\frac{227c}{1260}x^9+{\cal
O}\left(x^{11}\right), \quad x\to 0.
\end{eqnarray}
Let us now analyze the validity of the scaling regime. We ask that the
neglected terms in Eq.~(\ref{scaf}) remain small in the range $k_0\ll
k\ll \Lambda$. Using the large $x$ asymptotic for $f(x)$, we find that
the non linear term in $f$ appearing in Eq.~(\ref{scaf}) can be
neglected, if
\begin{equation}
k_0^{2\alpha-3}\frac{f'^2}{x}\ll
k_0^{\alpha-2}\quad\Rightarrow\quad k\ll k_0^{-(\alpha-2)},
\end{equation}
implying $\alpha\geq 2$. Similarly, the term arising from the time
derivative of the residual density is negligible provided that
\begin{eqnarray}
\dot k_0 k_0^{\alpha-1}\left(\alpha f-x f'\right)\ll
k_0^{\alpha-2}\quad&\Rightarrow &\quad k\ll
k_0^{\frac{3-\alpha}{2}} \quad (\alpha\ne 2),\\
&\Rightarrow &\quad k\ll \Lambda \quad (\alpha= 2),
\end{eqnarray}
which implies that $\alpha\geq 3$ or $\alpha=2$. We will find in
the next subsections that the admissible values $\alpha=3$ and
$\alpha=2$ are in fact respectively associated to the collapse
dynamics at $T_c$ and strictly below $T_c$.

We would like to emphasize that the present model provides an
interesting example where one obtains a scaling solution which scales
differently from what would have been obtained from a naive power
counting, assuming that {\it all} terms in Eq.~(\ref{scaf}) scale the
same way (which would lead to $\alpha=1$).

\subsection{Collapse at $T=T_c$}

At $T_c$, the constant appearing in the scaling function is
denoted $c_c$. Expressing the conservation of mass, we find
\begin{eqnarray}
1-2T_c\left(\Lambda-k_0\arctan\left(\frac{\Lambda}{k_0}\right)\right)&=&
F(\Lambda,k_0)\sim k_0^\alpha
f\left(\frac{\Lambda}{k_0}\right),\\
\frac{\pi}{2}\Lambda^{-1} k_0 &\sim &\frac{\pi c_c}{6} \Lambda^2
k_0^{\alpha-2}+{\cal O}\left(k_0^{2}\right),
\end{eqnarray}
which implies that
\begin{equation}
\alpha=3,\quad c_c\sim \Lambda^{-3}\sim T_c^3.
\end{equation}
Inserting this value for $\alpha$ in Eq.~(\ref{dk0dt}), we find
that $k_0(t)$ decays exponentially at $T_c$, with a rate $c_c$. To
complete the computation of the scaling function, we need to
determine the constant $c_c$ appearing as a prefactor in
Eq.~(\ref{fprime}), and which also controls this exponential
decay. Although the functional form of the scaling functions are
identical for the SM and BEM, we will see that the constant $c_c$
will differs for both models, as this constant is not entirely
determined by the dynamics in the region of high density, where
the term $(\rho+1)$ can be safely replaced by $\rho$. To make this
point clearer, we now look for a global solution for the
correction term $F(k,k_0)$, valid for momentum much greater than
$k_0$, and up to $k=\Lambda$:
\begin{equation}
F(k,k_0)\approx k_0\phi(k),\quad k_0\ll k\leq \Lambda.
\end{equation}
Note that the large $x=\frac{k}{k_0}$ asymptotic of the scaling
function should match the small $k$ behavior of $\phi(k)$:
\begin{equation}
k_0\phi(k)\sim_{k\to 0}k_0^3f\left(\frac{k}{k_0}\right)\sim_{k\gg
k_0} \frac{\pi c_c}{6}k_0 k^2.\label{asym}
\end{equation}
For the SM, expanding the integrated density up to
leading order in $k_0$
\begin{equation}
M(k,t)= 2T_c\left(k-\frac{\pi}{2}k_0\right)+k_0\phi(k) +{\cal
O}\left(k_0^{2}\right),\label{map}
\end{equation}
we immediately find that 
\begin{equation}
\phi(\Lambda)=\pi T_c.\label{cond}
\end{equation}
On the other hand, substituting Eq. (\ref{map}) in Eq. (\ref{mod3}) we
obtain to leading order in $k_{0}$:
\begin{equation}
\frac{\partial M}{\partial
t}=k_0T_c\left(\phi''+\frac{2}{k}\phi'\right)=\dot k_0
(\phi(k)-\pi T_c)= -c_ck_0(\phi(k)-\pi T_c),
\end{equation}
so that $\phi(k)$ satisfies
\begin{equation}
\phi''+\frac{2}{k}\phi'-\beta_c{c_c}\left({\phi(\Lambda)}-{\phi}\right)=0.
\end{equation}
This equation admits the solution
\begin{equation}
\phi(k)=\pi T_c\left(1-\frac{\sin(\sqrt{\beta_c
c_c}\,k)}{\sqrt{\beta_c c_c}\,k}\right),
\end{equation}
and expressing the condition of Eq.~(\ref{cond}), we determine
$c_c$ and the full function $\phi(k)$
\begin{equation}
c_c=4\pi^2T_c^3, \quad \phi(k)=\pi T_c\left(1-\frac{\sin(\pi
k/\Lambda)}{\pi k/\Lambda}\right).\label{phisimp}
\end{equation}
It is straightforward to check that $\phi(k)$ obeys the small $k$
behavior of Eq.~(\ref{asym}), identical to the $k\gg k_0$ behavior
of the scaling function $f$.

\vskip 1.5cm
\begin{figure}[ht]
\includegraphics[width=11.5cm]{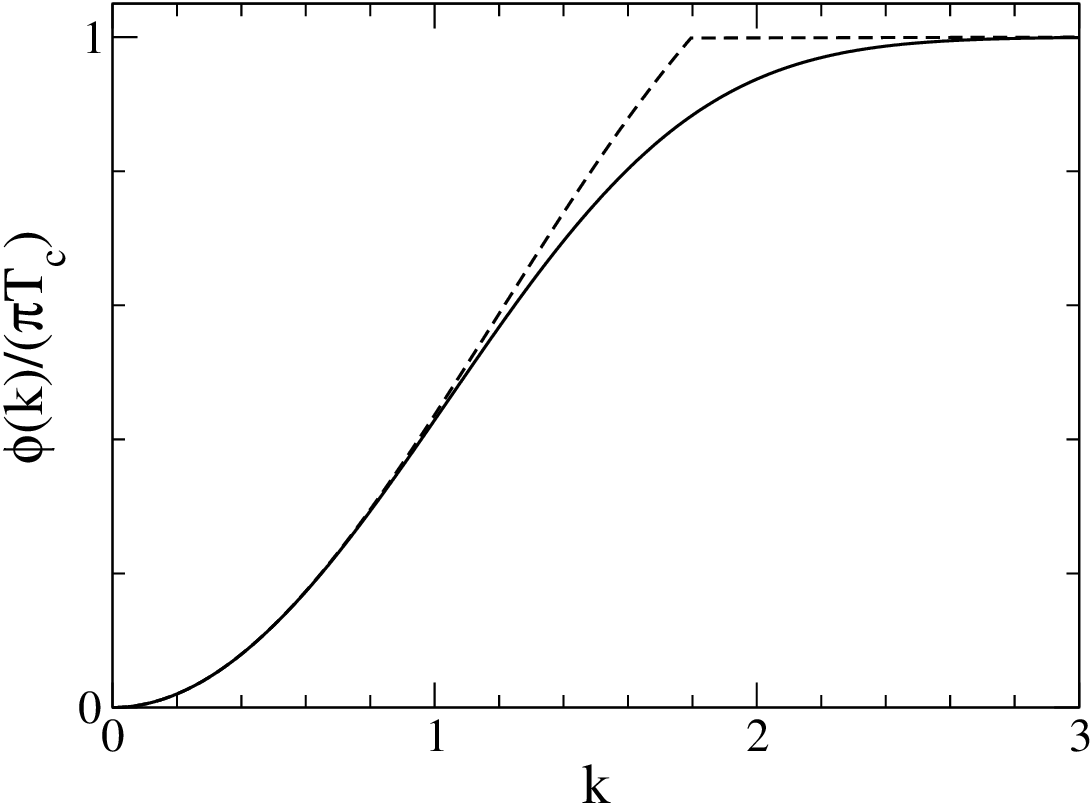}
\caption{We plot the analytical expressions of $\phi(k)$ (see
Eq.~(\ref{phisimp}), dashed line) for the simplified model
compared to the numerical solution of Eq.~(\ref{phireel}) (full
line). We have chosen the size of the confining box to be
$\Lambda=1.79805...$ so that the small $k$ behavior of the two
functions coincides.}\label{fig1}
\end{figure}
Finally, the collapse at $T_c$ for the SM has been fully analyzed,
and the central density (\ref{rhok}) diverges as
\begin{equation}
\rho(0,t)=2T_ck_0^{-2}(t)\sim\exp\left(8\pi^2T_c^3t\right).
\end{equation}
For infinite time, one converges to the expected equilibrium solution
\begin{equation}
M(k,\infty)=2T_c k,\quad \rho(k,\infty)=2T_c k^{-2}.
\end{equation}
\vskip 1.5cm
\begin{figure}[ht]
\includegraphics[width=11.5cm]{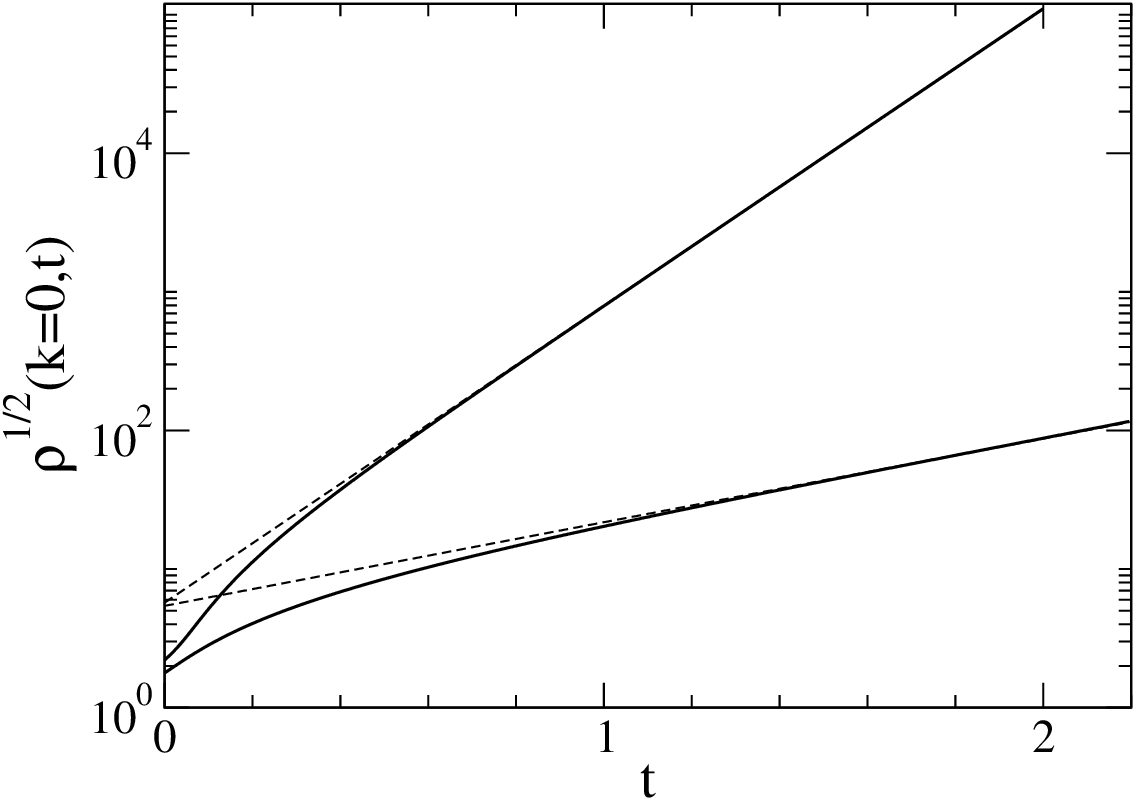}
%\vskip 0.9cm
\caption{At $T=T_c$, we plot $\rho^{1/2}(k=0,t)\sim
k_0^{-1}(t)\sim \exp(c_c t)$ computed numerically by integrating
the dynamical equation for $M(k,t)$ (see Eq.~(\ref{mod3d3})), for
the SM ($\Lambda=1$, top full line) and the BEM (bottom full
line). In both cases, we find an exponential growth, with a rate
in perfect agreement with our exact results for $c_c$ (the
straight dashed lines in these semi-log plots have a slope equal
to the theoretical values for $c_c$). For both models, we obtain
an excellent fit of $\rho^{1/2}(k=0,t)$ to the functional form
$\rho^{1/2}(k=0,t)=A\exp(c_c t)-B$, with $A\sim 5.7$ and $B\sim
0.5$ (fit not shown, but indistinguishable from data starting from
$t\sim 0.2$).}\label{fig2}
\end{figure}

\begin{figure}[ht]
\includegraphics[width=11.5cm]{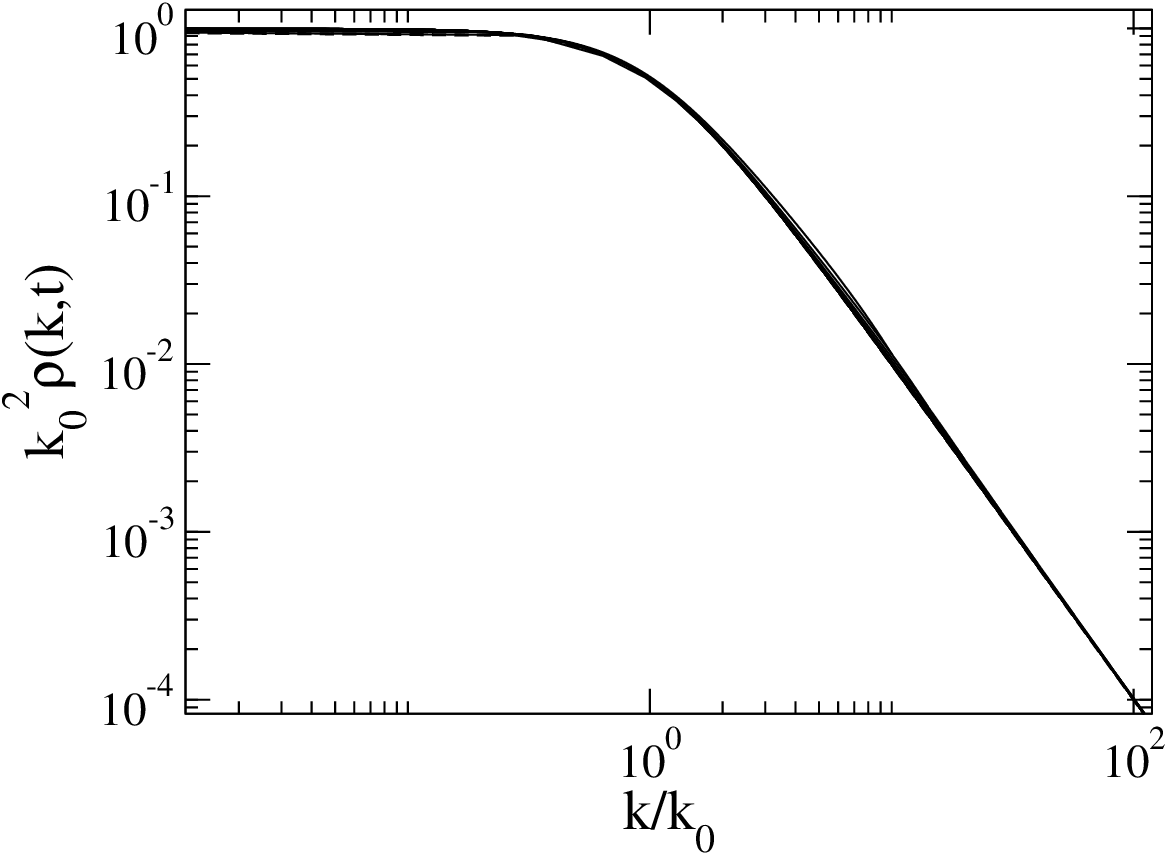}
%\vskip 0.9cm
\caption{At $T=T_c$, we plot $(2T_c)^{-1}k_0^2\rho(k,t)$ as a
function of $x=k/k_0$ for the BEM, for times for which the central
density $\rho(0,t)=2T_ck_0^{-2}$ is equal to $100{\times} 2^n$
($n=1,...,14$). We have also plotted the corresponding scaling
function $1/(1+x^2)$ (dashed line), which is perfectly
superimposed with the data collapse. A similar plot for the SM
would be indistinguishable, for the scale of momentum shown. Below
$T_c$, a similar scaling arises for both models.}\label{fig3}
\end{figure}

Now, this approach can be repeated for the BEM, starting with a
similar ansatz
\begin{equation}
M(k,t)=\int_{0}^{k}\frac{k'^{2}}{\exp\left(\beta_c\frac{
k'^2+k_0^2}{2}\right)-1}\,dk'+k_0\phi(k),\label{Mkt}
\end{equation}
where $\mu(t)=\beta_ck_0^2/2$ acts like a time-dependent chemical
potential. For $k\gg k_0$, we find that
\begin{equation}
\int_{0}^{k}\frac{k'^{2}}{\exp\left(\beta\frac{
k'^2+k_0^2}{2}\right)-1}\,dk'=\int_{0}^{k}
\frac{k'^{2}}{\exp\left(\beta\frac{ k'^2}{2}\right)-1}\,dk'-\pi T
k_0+{\cal O}\left(k_0^2\right).
\end{equation}
Using the above result, we can insert the expression of $M(k,t)$
from Eq.~(\ref{Mkt}) in the dynamical equation Eq.~(\ref{mod3d3}),
and find up to the leading order in $k_0$
\begin{equation}
\frac{\partial M}{\partial
t}=k_0T_c\left[\phi''+\left(\frac{\beta_c k}{
\tanh\left(\frac{\beta_c
k^2}{4}\right)}-\frac{2}{k}\right)\phi'\right]=\dot k_0
(\phi(k)-\pi T_c)= -c_ck_0(\phi(k)-\pi T_c),
\end{equation}
or, after simplification,
\begin{equation}
\phi''+\left(\frac{\beta_c k}{ \tanh\left(\frac{\beta_c
k^2}{4}\right)}-\frac{2}{k}\right)\phi'-\beta_c c_c
\left({\phi(\infty)}-{\phi}\right)=0,\label{phireel}
\end{equation}
with
\begin{equation}
\phi(\infty)=\pi T_c.
\end{equation}
Now, $c_c$ is selected by imposing that $\phi(k)$ is an increasing
function (being the integral of a density) and that one gets a
fast decay of $\phi(\infty)-\phi(k)\sim \exp(-\beta k^2/2)$.
Solving this eigenvalue problem numerically, we find
\begin{equation}
c_c=1.38452425...=C\pi^2 T_c^3,\quad C=1.50380614...
\end{equation}
which leads to
\begin{equation}
\rho(0,t)=2T_ck_0^{-2}(t)\sim\exp\left(2C\pi^2T_c^3t\right).
\end{equation}
In Fig.~\ref{fig1}, we have plotted the function $\phi(k)$ for the
BEM, compared to its analytical counterpart for the SM. As
expected, they significantly differ only for large momenta. In
Fig.~\ref{fig2}, we numerically confirm the exponential growth of
the density of particles at ${\bf k}={\bf 0}$ for both considered
models, with the growth rate $c_c$ in perfect agreement with the
present theoretical analysis. Finally, in Fig.~\ref{fig3}, we
illustrate the scaling behavior of the density profile.
\vskip 1.5cm

\subsection{Collapse for $0<T<T_c$}

We now repeat the above procedure below $T_c$. For the SM model,
expressing conservation of mass, we obtain
\begin{eqnarray}
1-2T\left(\Lambda-k_0\arctan\left(\frac{\Lambda}{k_0}\right)\right)&=&
F(\Lambda,k_0)\sim k_0^\alpha
f\left(\frac{\Lambda}{k_0}\right),\\
 \frac{T_c-T}{T_c}&\sim &\frac{\pi c}{6} \Lambda^2
k_0^{\alpha-2}+{\cal O}\left(k_0\right),
\end{eqnarray}
which implies that
\begin{equation}
\alpha=2, \quad c(T)\sim {T_c}{(T_c-T)}.
\end{equation}
For this value of $\alpha$, Eq.~(\ref{dk0dt}) can be integrated, leading to
\begin{equation}
k_0(t)=c(T)(t-t_{coll}).\label{k0stc}
\end{equation}
We thus find that, below $T_c$, a singularity should arise in a
finite time, for which the central density diverges when $t$ goes
to $t_{coll}$.

In order to obtain analytical results, we focus on the collapse
dynamics just below $T_c$, so that $\frac{T-T_c}{T_c}\ll 1$, hence
$F(k,k_0)\ll 1$ and $c\ll 1$. In this regime, non linear terms in
$F$ can be safely neglected. For momenta $k\gg k_0$, we define a
function $\phi(k)$ such that
\begin{equation}
F(k,k_0)\approx \left(1+\frac{\lambda}{c}k_0\right)\phi(k),\quad
k_0\ll k\leq \Lambda,
\end{equation}
resulting in the following expression for the integrated mass
density
\begin{equation}
M(k,t)= 2T\left(k-\frac{\pi}{2}k_0\right)+
\left(1+\frac{\lambda}{c}k_0\right)\phi(k)+{\cal
O}\left(k_0^{2}\right).
\end{equation}
Taking $k=\Lambda$ and matching terms of order ${\cal
O}\left(k_0^{0}\right)$ and ${\cal O}\left(k_0^{1}\right)$, we deduce
that
\begin{equation}
\phi(\Lambda)=\frac{T_c-T}{T_c}=\pi
\frac{c}{\beta\lambda}.\label{phiL}
\end{equation}
Inserting the expression of $M(k,t)$ in the dynamical equation
Eq.~(\ref{mod3d3}), we obtain
\begin{equation}
\frac{\partial M}{\partial
t}=T\left(\phi''+\frac{2}{k}\phi'\right)=\pi c\,T-{\lambda}\,
\phi,
\end{equation}
or
\begin{equation}
\phi''+\frac{2}{k}\phi'-
\beta\lambda\left({\phi(\Lambda)}-{\phi}\right)=0.\label{define}
\end{equation}
We recognize the very same equation for $\phi(k)$ as in the
preceding section, so that the two functions are identical up to a
multiplicative constant. Hence, for the SM, we obtain
\begin{equation}
\lambda=c_c=4\pi^2T_c^3, \quad c(T)=4\pi T_c(T_c-T), \quad
\phi(k)=\frac{T_c-T}{T_c}\left(1-\frac{\sin(\pi k/\Lambda)}{\pi
k/\Lambda}\right),\label{rest1}
\end{equation}
and a power law divergence of the central density
\begin{equation}
\rho(0,t)=2Tk_0^{-2}(t)\sim\frac{1}{8\pi^2
T_c^3}\left(\frac{T_c-T}{T_c}\right)^{-2}\left(t_{coll}-t\right)^{-2}.
\end{equation}

For the SM, we can evaluate $\phi(k)$, $c(T)$, and $\lambda(T)$ at the
next order in $\delta=\frac{T_c-T}{T_c}$. We write
\begin{eqnarray}
\phi(k) &=& \delta\, \phi_0(k) + \delta^2\,\phi_1(k), \\
c(T) &=& \delta \, c_0 + \delta^2 \,c_1, \\
\lambda(T) &=& \lambda_0 + \delta\,\lambda_1,
\end{eqnarray}
where $\phi_0(k)$, $c_0$ and $\lambda_0=c_c$ can be easily determined
from Eq.~(\ref{rest1}). Inserting the above ansatz in the dynamical
equation leads to a complicated linear equation for
$\phi_1(k)$. We must impose the boundary conditions
$\phi_1(0)=\phi_1'(0)=0$ and $\phi_1(\Lambda)=0$, which fix the value
of $c_1$ and $\lambda_1$. After some cumbersome but straightforward
calculations, we obtain
\begin{eqnarray}
\phi_1\left(x=\pi\frac{k}{\Lambda}\right) &=& \frac{\sin
x}x\left[\frac{c_1}{8 \pi T_c^2}\left(1 + \frac{\pi -
x}{\tan(x)}\right) + \frac{2 \pi \cos(x)}{3 x}\right. \\ && + \frac{3
\pi}{4 \tan(x)}\left(\int_{\pi}^{x} \frac{\cos(t)}{t}\,dt - \int_{3
\pi}^{3 x}\frac{\cos(t)}{t}\,dt\right) \nonumber \\ && +
\frac{\pi}{4}\left(3 \int_{3 x}^{+\infty}\frac{\sin(t)}{t} \,dt -
\frac{11}{3}\int_{x}^{+\infty}\frac{\sin(t)}{t}\,dt\right) \nonumber
\\ &&\left. + \frac{\pi (\cos(2 x)-1)}{6 x^2 \sin(x)}\right],\nonumber
\end{eqnarray}
with
\begin{eqnarray}
c_1 &=& \frac{2\pi T_c^2}{3} \left(9\ln(3)-4-9\int_\pi^{3\pi}
\frac{\cos(t)}{t}\,dt\right), \\
&=& C_1 T_c^2,\quad C_1= 13.51919467982...
\end{eqnarray}
and
\begin{equation}
\lambda_1=\pi C_1 T_c^3.
\end{equation}
Finally, we obtain
\begin{eqnarray}
c(T) &=& 4 \pi T_c (T_c - T) +C_1 (T_c - T)^2 +
\mathcal{O}\left((T-T_c)^3\right),\label{c1} \\
\lambda(T) &=& 4 \pi T_c^3 + \pi C_1 T_c^2 (T_c - T) +
\mathcal{O}\left(T_c(T-T_c)^2\right),
\end{eqnarray}
and note that the next correction to $c(T)$ and $\lambda(T)$ are both
positive.
\vskip 1.5cm
\begin{figure}[ht]
\includegraphics[width=11.5cm]{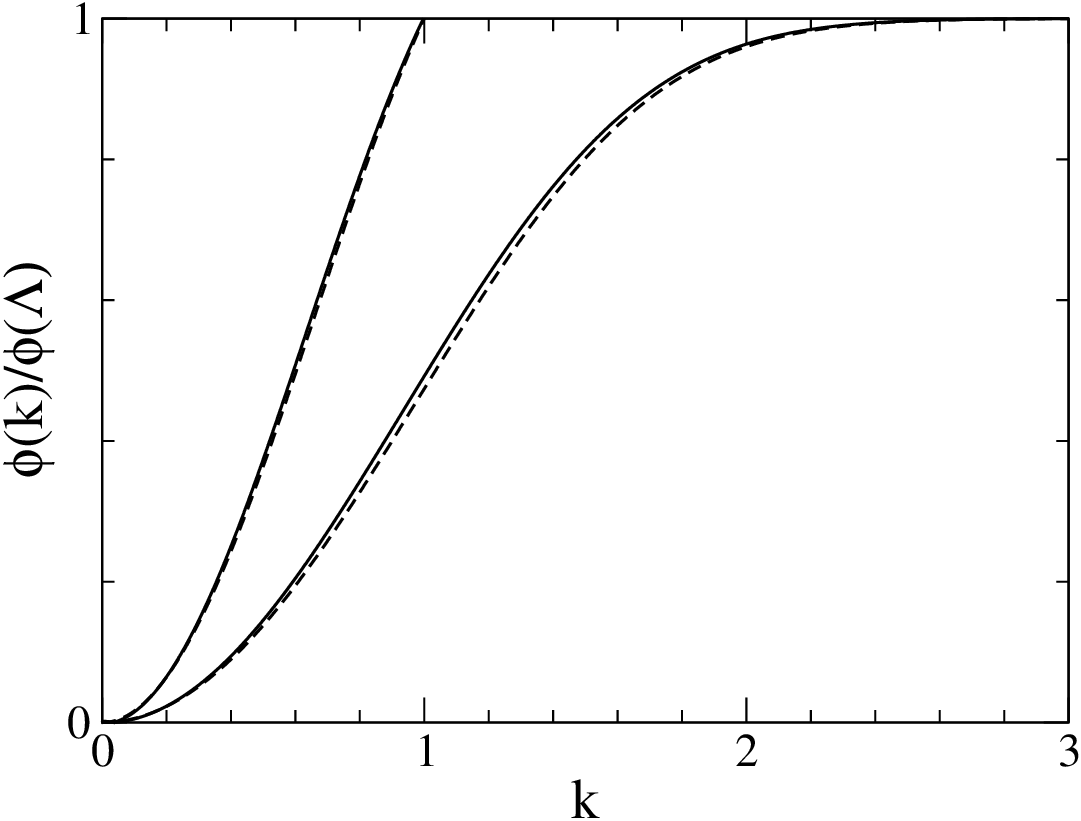}
%\vskip 0.9cm
\caption{For the simplified model, we plot
$\phi(k)=M(k,t_{coll})-2Tk$ obtained by numerically integrating
the dynamical equation for $M(k,t)$ at $T=0.9T_c$ (top full line).
We have chosen $\Lambda=1$ so that $T_c=1/2$ and
$\phi(\Lambda)=\frac{T_c-T}{T_c}=1/10$ (see Eq.~(\ref{phiL})). It
is in good agreement with our analytical expression of
Eq.~(\ref{rest1}) (top dashed line), which is strictly valid only
very close to $T_c$. We also plot the numerical (bottom full line)
and theoretical expression of $\phi(k)$ (bottom dashed line), for
the Bose-Einstein model
($\phi(\Lambda=\infty)=1-27/10^{3/2}=0.1462...\approx
\frac{3}{2}\frac{T_c-T}{T_c}$).}\label{fig4}
\end{figure}

Again, this approach can be repeated for the more realistic BEM,
by writing
\begin{equation}
M(k,t)=\int_{0}^{k}\frac{k'^{2}}{\exp\left(\beta\frac{
k'^2+k_0^2}{2}\right)-1}\,dk'+
\left(1+\frac{\lambda}{c}k_0\right)\phi(k).\label{mktsous}
\end{equation}
At the leading order in $k_0$, the dynamical equation
Eq.~(\ref{mod3d3}) leads to
\begin{equation}
\frac{\partial M}{\partial t}=T\left[\phi''+\left(\frac{\beta k}{
\tanh\left(\frac{\beta
k^2}{4}\right)}-\frac{2}{k}\right)\phi'\right]=\pi c\, T-
{\lambda}\, \phi,
\end{equation}
or more simply to
\begin{equation}
\phi''+\left(\frac{\beta k}{ \tanh\left(\frac{\beta
k^2}{4}\right)}-\frac{2}{k}\right)\phi'-
\beta\lambda\left({\phi(\infty)}-{\phi}\right)=0,
\end{equation}
which is again the same equation as the one found in the previous
section, which determines the function $\phi(k)$ up to a
multiplicative constant. In order to compute this constant, we
have to match the terms of order ${\cal O}\left(k_0^{0}\right)$
and ${\cal O}\left(k_0^{1}\right)$ for $k\to+\infty$ in
Eq.~(\ref{mktsous}), which implies that
\begin{eqnarray}
\phi(\infty)&=&1-\left(\frac{\beta_c}{\beta}\right)^{3/2}\approx
\frac{3}{2}\frac{T_c-T}{T_c},\\
&=&\pi \frac{c}{\beta\lambda}.
\end{eqnarray}
Finally, we obtain
\begin{equation}
\lambda=c_c=C\pi^2 T_c^3, \quad c(T)=C'\pi T_c(T_c-T),\quad
C'=\frac{3C}{2}=2.2557092066...
\end{equation}
and
\begin{equation}
\rho(0,t)=2Tk_0^{-2}(t)\sim\frac{2}{C'^2\pi^2
T_c^3}\left(\frac{T_c-T}{T_c}\right)^{-2}\left(t_{coll}-t\right)^{-2}.
\label{div}
\end{equation}

\vskip 1.5cm
\begin{figure}[ht]
\includegraphics[width=11.5cm]{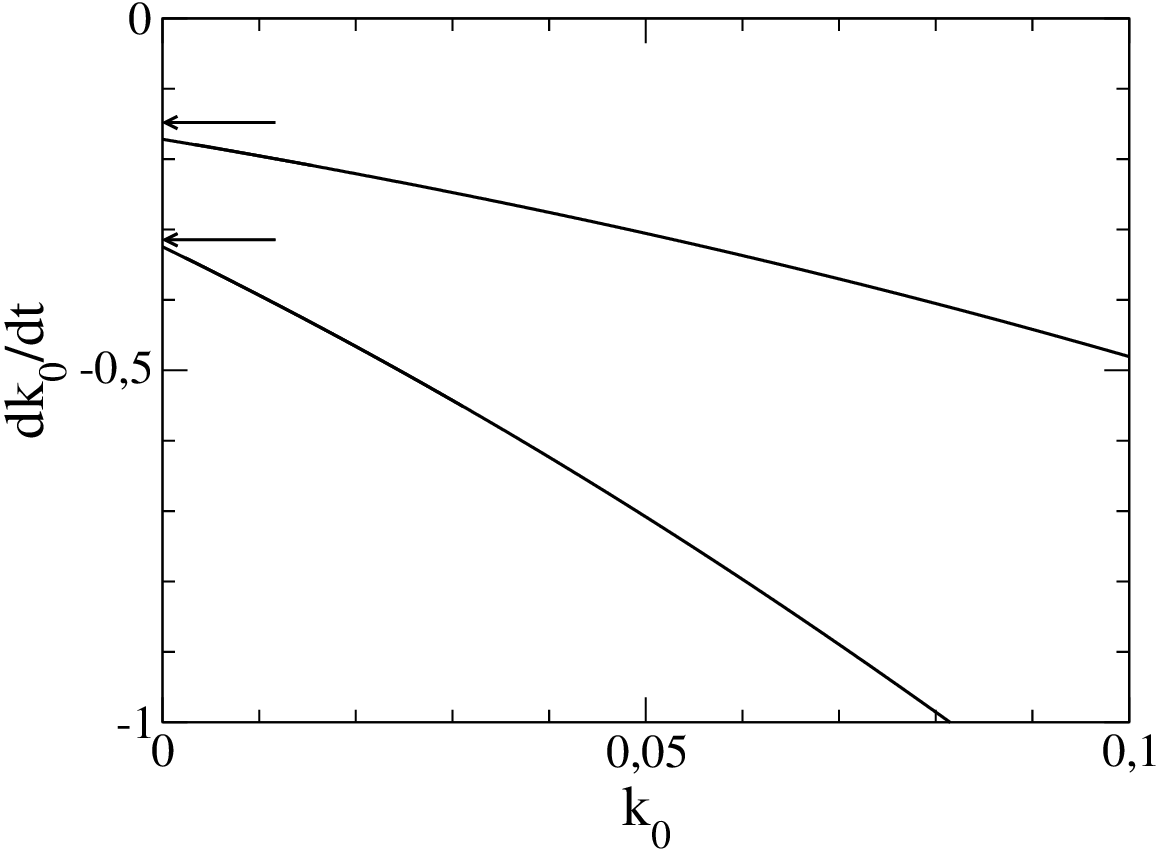}
%\vskip 0.9cm
\caption{For the simplified model ($\Lambda=1$, bottom line) and
the Bose-Einstein model (top line), we plot $\dot k_0(t)\to -c(T)$
as a function of $k_0(t)$ as obtained by integrating the dynamical
equation for $M(k,t)$ at $T=0.9T_c$. The theoretical slopes
$-c(T)$ evaluated at the first order in $T_c-T$ are indicated by
an arrow. We find that the numerical value of $c(T)$ are slightly
bigger than our first order calculation, consistent with the fact
that the next correction is analytically found to be positive for
the SM (see Eq.~(\ref{c1})). For small $k_0(t)$ ({\it i.e.} close
to $t_{coll}$), we find $\dot k_0(t)\approx -c(T)-A c_c\, k_0(t)$,
with $A\approx 1.35$ for the SM, and $A\approx 1.65$ for the
BEM.}\label{fig5}
\end{figure}

Note that at $t=t_{coll}$, we find that the density and the
integrated density are
\begin{equation}
\rho(k,t_{coll})=\frac{2T}{k^2}+\frac{\phi'(k)}{k^2},\quad
M(k,t_{coll})=2Tk+\phi(k),
\end{equation}
for the simplified model, and
\begin{equation}
\rho(k,t_{coll})= \frac{1}{\exp\left(\beta\frac{
k^2}{2}\right)-1}+\frac{\phi'(k)}{k^2},\quad
M(k,t_{coll})=\int_{0}^{k}\frac{k'^{2}}{\exp\left(\beta\frac{
k'^2}{2}\right)-1}\,dk'+\phi(k),
\end{equation}
for the more realistic Bose-Einstein model. In addition to the
equilibrium density, we obtain in both models a singular
contribution near $k=0$
\begin{equation}
\frac{\phi'(k)}{k^2}\sim \frac{\pi }{3k}c(T),\quad k\to 0.
\end{equation}
However, we do not observe the appearance of a Dirac peak at ${\bf
k}={\bf 0}$, implying that the evolution of the system is not
finished yet. Hence, we expect that after $t_{coll}$, the mass
included in $\phi(k)$ will be swallowed at $k=0$, finally giving
rise to the condensate.

Before addressing this issue, we give an estimate of $t_{coll}$,
near $T_c$. If one is very close to $T_c$, the density first grows
exponentially like at $T_c$, before crossing over to the behavior
of Eq.~(\ref{div}), when the time is of order $t_{coll}$. Matching
the two regimes, we obtain
\begin{equation}
\rho(k=0,t\sim t_{coll})\sim\exp\left(2c_ct_{coll}\right)\sim
\left(\frac{T_c-T}{T_c}\right)^{-2}t_{coll}^{-2},
\end{equation}
leading to the rough estimate
\begin{equation}
t_{coll}\sim T_c^{-3}\ln\left(\frac{T_c}{T_c-T}\right).\label{tcolldef}
\end{equation}
As expected, $t_{coll}$ diverges as the temperature approaches
$T_c$. This analysis suggests a global form for $\dot k_0(t)$,
valid for all temperature
\begin{equation}
\dot k_0=-c(T) h\left(\frac{c_c}{c(T)}k_0\right),\quad
h(0)=1,\quad h(x)\sim_\infty x,\label{hhh}
\end{equation}
where $c(T)\sim T_c(T_c-T)$ for both models. If we assume the simple
form $h(x)=1+x$ (numerically we find $h(x)=1+A x$, for
small $x$; see Fig.~\ref{fig5}), which is compatible with the known
asymptotics of $f(x)$ in Eq.~(\ref{hhh}), we obtain
\begin{equation}
\dot k_0=-c_c k_0-c(T),
\end{equation}
This equation has a global solution
\begin{equation}
k_0(t)=\frac{c(T)}{c_c}
\left(\exp\left[c_c(t_{coll}-t)\right]-1\right). \label{k0full}
\end{equation}
For times $t$ close to $t_{coll}$, this expression leads to the
asymptotic scaling of Eq.~(\ref{k0stc}), whereas it also
reproduces the early exponential decay with rate $c_c$, which is
expected for $T$ close to $T_c$. Imposing that at $t=0$, $k_0$
should be of order of a typical momentum of the initial condition
(for instance $k_0\sim \Lambda$ for the SM), we recover exactly
the estimate of $t_{coll}$ obtained in Eq.~(\ref{tcolldef}).

In Fig.~\ref{fig4}, we plot the numerical estimate of $\phi(k)$
obtained by numerically integrating Eq.~(\ref{mod3d3}) at
$T=0.9T_c$, and compare it to our theoretical results, which are,
in principle, only valid very close to $T_c$. In Fig.~\ref{fig5},
we plot  $\dot k_0(t)\to -c(T)$ obtained by numerically integrating
Eq.~(\ref{mod3d3}) and find a fair agreement with our theoretical
estimates for $c(T)$, which are strictly valid only close to $T_c$.

\subsection{Collapse at $T=0$}

For completeness, we now consider the collapse dynamics at $T=0$.
Since we are interested in the density scaling function, we
address this question within the simplified model. We consider the
dynamical equation for the density rather than for the integrated
density
\begin{equation}
\frac{\partial \rho}{\partial t}=3\rho^2+2k\rho\rho'.
\label{modt0}
\end{equation}
Inserting a scaling ansatz of the form
\begin{equation}
\rho(k,t)=\rho_0 g\left(\frac{k}{k_0}\right),
\end{equation}
in the dynamical equation Eq.~(\ref{modt0}), we obtain
\begin{equation}
\dot\rho_0 g-\rho_0\frac{\dot k_0}{k_0}x
g'=\rho_0^2\left(3g^2+2xgg'\right).
\end{equation}
We introduce the parameter $\alpha$ defined by
\begin{equation}
\frac{\dot\rho_0}{\rho_0}=-\alpha\frac{\dot
k_0}{k_0}=\alpha\rho_0.
\end{equation}
Then, the equation for the scaling function becomes
\begin{equation}
\frac{1-2g}{g(\alpha -3g)}g'=-\frac{1}{x},
\end{equation}
which can be exactly solved, leading to the following implicit
equation for $g(x)$
\begin{equation}
g(x)\left(
\frac{\alpha}{3}-g(x)\right)^{2\alpha/3-1}=Cx^{-\alpha},\label{scat0}
\end{equation}
where $C$ is a constant depending on the initial conditions, as
was already noticed within the study of the gravitational collapse
at $T=0$ \cite{charosi,csdim,cspostcoll}. Now, for small $x=k/k_0$, $g(x)$
must be an analytic function, with a small $x$ expansion of the
form $g(x)=g(0)+g''(0)x^2/2+...$. Matching the small $x$ behavior
of the RHS and LHS of Eq.~(\ref{scat0}), we obtain
\begin{equation}
\alpha=\frac{6}{7},\quad g(0)=\frac{2}{7},
\end{equation}
and
\begin{equation}
g(x)\left( \frac{2}{7}-g(x)\right)^{-3/7}=Cx^{-6/7}.
\end{equation}
We also obtain the exact expression for the central density, and
the time-dependent width of the dense core $k_0(t)$
\begin{equation}
\rho(k=0,t)=\frac{2}{7}\,\rho_0(t)=\frac{1}{3}(t_{coll}-t)^{-1},\quad
k_0(t)=(t_{coll}-t)^{7/6}.
\end{equation}
Finally, at $t=t_{coll}$, the small $k$ behavior of the density
becomes universal
\begin{equation}
M(k,t_{coll})\sim k^{15/7},\quad \rho(k,t_{coll})\sim k^{-6/7},
\end{equation}
in contrast with the behavior obtained for $0<T\leq T_c$, where we
found $\rho(k,t_{coll})\sim 2Tk^{-2}$.

\section{Post-collapse dynamics for $0<T<T_c$}

As mentioned in Section III.C, the density profile at $t=t_{coll}$
is not yet equal to the equilibrium profile which presents a Dirac
peak at ${\bf k}={\bf 0}$. This means that after $t_{coll}$, the
density profile should continue to relax to
\begin{equation}
\rho({\bf k})=\frac{T_c-T}{T_c}\delta({\bf k})+2Tk^{-2},
\end{equation}
for the simplified model, and
\begin{equation}
\rho({\bf
k})=\left[1-\left(\frac{\beta_c}{\beta}\right)^{3/2}\right]\delta({\bf
k})+ \frac{1}{\exp\left(\beta\frac{k^2}{2}\right)-1},
\end{equation}
for the Bose-Einstein model. In order to obtain analytical results
in this post-collapse stage, we again consider the case of a
temperature just below $T_c$, such that
\begin{equation}
\frac{T_c-T}{T_c}\ll 1.
\end{equation}
In this regime, and for the SM, the following ansatz is an exact
solution of the dynamical equation Eq.~(\ref{mod3d3})
\begin{equation}
M(k,t)=M_0(t)+2Tk+\left(1-\frac{M_0(t)}{\phi(\Lambda)}\right)\phi(k),
\end{equation}
where $M_0(t)$ is the time-dependent mass in the condensate, and
where $\phi(k)$ has been calculated in the collapse regime in the
previous section. Inserting this ansatz in the kinetic equation,
we obtain
\begin{equation}
\frac{\partial M}{\partial t}=\dot M_0
\left(1-\frac{\phi(k)}{\phi(\Lambda)}\right)=
T\left(1-\frac{M_0(t)}{\phi(\Lambda)}\right)
\left(\phi''+\frac{2}{k}\phi'\right),
\end{equation}
where we have neglected quadratic terms in $\phi(k)$, which is
justified near $T_c$. For this equation to be compatible with
Eq.~(\ref{define}), {\it i.e.} the defining equation for
$\phi(k)$, we must have
\begin{equation}
\dot M_0=c_c \left(\phi(\Lambda)-{M_0(t)}\right), \label{dotm0}
\end{equation}
which can be easily solved, leading to the full time dependence of
the condensate mass $M_0(t)$
\begin{equation}
M_0(t)=\frac{T_c-T}{T_c}
\left(1-\exp\left[-4\pi^2T_c^3(t-t_{coll})\right]\right).
\end{equation}
For the Bose-Einstein model, we proceed in a similar manner, and
find for $T$ close to $T_c$
\begin{eqnarray}
M_0(t)&=&\left[1-\left(\frac{\beta_c}{\beta}\right)^{3/2}\right]
\left(1-\exp\left[-C
\pi^2T_c^3(t-t_{coll})\right]\right),\\
&\approx &
\frac{3}{2}\frac{T_c-T}{T_c} \left(1-\exp\left[-C
\pi^2T_c^3(t-t_{coll})\right]\right).
\end{eqnarray}
For both models, the condensate mass initially grows linearly with
time
\begin{equation}
M_0(t)=\pi T_c c(T)(t-t_{coll})\sim T_c^2(T_c-T)(t-t_{coll}),\quad
t-t_{coll}\ll T_c^{-3},
\end{equation}
whereas it saturates exponentially fast to its equilibrium value,
with a rate $c_c$ equal to the relaxation rate found at $T_c$ (see
Section III.B and the next section), since one can write
\begin{equation}
1-\frac{M_0(t)}{M_0(\infty)}= \exp\left[-c_c(t-t_{coll})\right].
\end{equation}

\section{Relaxation time for $T>T_c$}

Finally, in this section, we address the problem of determining
the relaxation rate to the equilibrium solution above $T_c$.
Writing
\begin{equation}
M(k,t)=\int_{0}^{k}\frac{k'^{2}}{\exp\left(\frac{ \beta
k'^2}{2}+\mu\right)-1}\,dk'+\epsilon(k)\exp(-t/\tau),
\end{equation}
and neglecting non linear terms in $\epsilon(k)$, we find that
\begin{equation}
\epsilon(k)= h(\sqrt{\beta}k),
\end{equation}
where $h(x)$ satisfies
\begin{equation}
h''+\left(\frac{x}{
\tanh\left(\frac{x^2}{4}+\frac{\mu}{2}\right)}-\frac{2}{x}\right)
h' +\tau^{-1}h=0.\label{eigen}
\end{equation}
For $T\gg T_c$ (and hence $\mu\gg 1$), we can replace $\tanh$ by unity
and  the solution can be exactly
written in term of the Kummer confluent hypergeometric function
$K$
\begin{equation}
h(x)=x^3 K\left(\frac{3\tau+1}{2\tau},
\frac{5}{2},-\frac{x^2}{2}\right).
\end{equation}
Imposing a fast decaying solution, we find
\begin{equation}
h(x)=x^3\exp\left(-\frac{ x^2}{2}\right),\quad
\tau(T\to+\infty)=\frac{1}{2}.
\end{equation}
For the simplified model, the limit of high temperature is not
physically interesting since, due to the finite box, the particles
cannot spread up to large momenta.

For $0<T-T_c\ll T_c$ (and hence $\mu\ll 1$), the eigenequation
Eq.~(\ref{eigen}) coincide with the defining equation for the
function $\phi(k)$ introduced in Section III.B. Hence, $\tau(T\to
T_c)=c_c^{-1}$, using the respective values of $c_c$ for the two
models considered.

It seems paradoxical that the relaxation time does not diverge at
$T_c$, contrary to what is expected for a continuous phase
transition. However, $\tau$ is {\it not} the equilibration time, which
is the typical time $\tau_{eq}$ needed to reach the equilibrium
solution. This time can be evaluated by considering that close but
above $T_c$, the chemical potential initially decreases as $\beta
k_0(t)^2/2$ does at $T_c$, and reaches the order of magnitude of its
equilibrium value after a time $\tau_{eq}$. Thus, we write
\begin{equation}
k_0(\tau_{eq},T_c)\sim\exp(-c_c \tau_{eq})\sim k_0(T)=\sqrt{2\mu
T},
\end{equation}
Using the expressions for $\mu$ near $T_c$ given in
Eq.~(\ref{mu1}) for the BEM and in Eq.~(\ref{mu2}) for the SM, we
obtain
\begin{equation}
\tau_{eq}\sim - T_c^{-3}\ln\mu\sim
T_c^{-3}\ln\left(\frac{T_c}{T-T_c}\right),
\end{equation}
for both models. Note the similarity of this estimate with the
expression of $t_{coll}$ near $T_c$, given in
Eq.~(\ref{tcolldef}). The exponential relaxation controlled by
$\tau$ only starts after a time of order $\tau_{eq}\gg\tau$.

\section{Comparison with other works}
\label{sec_comp}

In this section, we give a short review of classical works concerning
the dynamics of the Bose-Einstein condensation to show how our results
compare with these studies. The dynamics of the Bose-Einstein
condensation has been described by two apparently different kinetic
theories. The first kinetic theory is based on a quantum version of
the Boltzmann equation for the one-particle distribution function
$\rho({\bf k},t)$. This is a semi-classical approach which introduces
corrections for quantum statistics into the ordinary Boltzmann
collision term \cite{nordheim,bloch,uu}. Another description is provided by
the time-dependent Gross-Pitaevskii equation for the condensate wave
function $\psi({\bf x},t)$ (order parameter for the Bose condensate)
in a spatially homogeneous medium \cite{gp}. This equation does not take into
account quantum fluctuations, or thermal or irreversible effects, but
is valid when a large number of particles have condensed.

One of the first solution of these kinetic equations is due to Levich
\& Yakhot \cite{ly} who consider a gas of bosons without interaction
(i.e. neglecting collisions) in contact with a thermal bath of
fermions. They provide an analytical solution of the corresponding quantum
Boltzmann equation and find that the Bose condensate forms in an
infinite time. Since this result is inconsistent with observations,
Stoof \cite{stoof1} argues that the quantum Boltzmann equation is not valid to
describe the condensate. He considers an isolated gas of bosons in
interaction (i.e. with interatomic collisions) and argues that the
dynamics of the collapse follows three steps: (i) an incoherent
evolution described by the quantum Boltzmann equation (ii) a coherent
evolution triggering a phase transition (instability) leading to the
Bose condensate (iii) a thermalization between the condensate and the
non condensed atoms interpreted as quasi-particles in the sense of
Bogoliubov. Stoof focuses on the coherent evolution. Using a
functional formulation of the Keldysh theory, he derives a time
dependent Landau-Ginzburg equation for the order parameter $\langle
\psi({\bf x},t)\rangle$ of the phase transition. He also shows that
the transition temperature for interacting bosons is larger than for
an ideal gas and that the nucleation of the condensation is short,
contrary to the result of \cite{ly}, except for temperatures close to
the critical temperature $T_{c}$.  In a more recent paper, Stoof
\cite{stoof2} derives a Fokker-Planck equation for the probability
distribution of the order parameter $\langle
\psi({\bf x},t)\rangle$ which gives a unified description of 
both the incoherent (kinetic Boltzmann) and coherent
(Gross-Pitaevskii) stages of Bose-Einstein condensation. A similar
program of unification of the two theories is carried out by Gardiner
\& Zoller \cite{gz}. Using a projection of the density operator $\rho$, they
derive a quantum kinetic master equation (QKME) for bosonic atoms and
recover, in particular limits, the quantum Boltzmann equation and the
Gross-Pitaevskii equation.

Semikoz \& Tkachev \cite{st} and Lacaze {\it et al.} \cite{lacaze}
numerically solve the quantum Boltzmann equation and find the
formation of a condensate in finite time and the growth of this
condensate in a post-collapse regime. This is quite different from the
results of Levich \& Yakhot \cite{ly} which are valid when the system
is coupled to a bath of fermions. Therefore, the original bosonic Boltzmann
kinetic equation (without fermion bath approximation) can account for
a finite time formation of the condensate. More recently, using an
analogy with optical turbulence, Connaughton \& Pomeau \cite{cp}
obtain a kinetic equation for the spectral particle density $\rho({\bf
k},t)$ directly from the Gross-Pitaevskii equation for the wave
function $\psi({\bf x},t)$ when the non-linear term is considered as a
perturbation (${\bf k}$ is the conjugate of ${\bf x}$ in the Fourier
analysis). They show that this kinetic equation has the same form as
the quantum Boltzmann equation.

Several authors \cite{svistunov,kagan,st,lacaze} have investigated
self-similar solutions of the quantum Boltzmann equation in the form
$\rho(\epsilon,t)=\rho_{0}(t)f(\epsilon/\epsilon_{0}(t))$, where
$\epsilon=\frac{k^2}{2m}$. In particular, Semikoz \& Tkachev \cite{st}
find a pre-collapse regime generating, in a finite time $t_{coll}$, a
distribution function of energies scaling as $f(\epsilon)\sim
\epsilon^{-\alpha}$ with $\alpha=1.24$. This profile is slightly
steeper than the Zakharov profile $\epsilon^{-7/6}$ providing an exact
static solution of the quantum Boltzmann equation corresponding to a
constant mass flux $J$ in momentum space toward the condensate. The
central density increases as $\rho_{0}(t)\sim
(t_{coll}-t)^{-\alpha/2(\alpha-1)}\sim (t_{coll}-t)^{-2.6}$ and
becomes infinite at $t=t_{coll}$. The typical core energy
$\epsilon_{0}(t)\sim (t_{coll}-t)^{1/2(\alpha-1)}\sim
(t_{coll}-t)^{2.1}$ goes to zero at $t=t_{coll}$.  Semikoz \& Tkachev
\cite{st} also investigate the post-collapse regime. They find that
the energy distribution passes from $\epsilon^{-\alpha}$ to
$\epsilon^{-1}$ after the singularity and that the mass of the
condensate grows like
$n_{0}(t)=(t-t_{coll})^{(3-2\alpha)/4(\alpha-1)}\sim
(t-t_{coll})^{0.54}$ just after collapse.  Lacaze {\it et al.}
\cite{lacaze} also investigate self-similar solutions and obtain
similar results with an exponent $\alpha=1.234$. 

Our approach is physically different since we consider a gas of
non-interacting bosons in contact with a heat bath establishing a
Bose-Einstein distribution (canonical description) while the previous
authors consider an isolated system of bosons in interaction
(microcanonical description). However, although the dynamics differs
in the details (as expected), the phenomenology of the collapse is almost the same and the
simplified form of our kinetic equation (\ref{bose1}) allows for a
complete analytical solution of the problem (pre and post collapse)
which is not possible for the quantum Boltzmann equation
\cite{svistunov,kagan,st,lacaze}. This is clearly an interest of our model.

\section{Conclusion}

In the present paper, we have considered the condensation in ${\bf
k}$-space of a homogeneous gas of non-interacting bosons in a thermal
bath fixing the temperature $T$. In statistical mechanics, this
corresponds to a {\it canonical} description.  We have pointed out the
striking analogy between the dynamical equation describing the
Bose-Einstein condensation in the canonical ensemble and the one
describing the evolution of a self-gravitating gas of Brownian
particles (or the chemotaxis of bacterial populations).

In analogy to the case of a classical gas, we have constructed the
dynamical Fokker-Planck equation for the momentum distribution.  For
$T<T_{c}$, the system experiences a finite time singularity, {\it
i.e.} the density at ${\bf k}={\bf 0}$ becomes infinite in a finite
time $t_{coll}\sim
T_c^{-3}\ln\left(\frac{T_c}{T-T_c}\right)$. However, the ``singularity
contains no mass'' and the condensate is formed during the
post-collapse regime for $t>t_{coll}$.  This dynamical scenario leads,
for $t\rightarrow +\infty$, to a Dirac peak in addition to the
Bose-Einstein distribution with zero chemical potential, which is
entirely consistent with the distribution predicted by Einstein at
equilibrium \cite{einstein}. Our canonical description permits a
complete analytical treatment for the different stages of the process,
contrary to the microcanonical approach starting from the
semi-classical Boltzmann equation \cite{st,lacaze}, where one has to
mostly rely on numerical simulations. Although qualitatively similar
(existence of a finite time pre-collapse regime for $t<t_{coll}$,
scaling behavior of the density, formation of the condensate in the
post-collapse regime...), we note that our results differ
quantitatively from the ones obtained within the microcanonical
approach. For instance, $k_0(t)=c(T)(t_{coll}-t)$ in our canonical
model, whereas $k_0(t)\sim (t_{coll}-t)^{\gamma}$ in the
microcanonical ensemble, where, numerically, one finds $\gamma\approx
1.07$ (the difference is even more apparent on other quantities)
\cite{lacaze}.  This should not be surprising, as the same phenomenon
arises in the study of the collapse dynamics of a self-gravitating
gas, which is different in the canonical and microcanonical ensembles,
though qualitatively similar in many respects \cite{charosi}.

In this paper, we have assumed that the Bose gas is homogeneous.
The inhomogeneous situation could be studied semiclassically by
introducing a mean-field advective term in the kinetic equation
such that
\begin{equation}
\frac{\partial\rho}{\partial t}+{\bf k}\cdot \frac{\partial
\rho}{\partial {\bf r}}-\nabla U({\bf r})\cdot \frac{\partial
\rho}{\partial {\bf k}}=Q(\rho),
\end{equation}
where $Q(\rho)$ is either the bosonic Fokker-Planck operator of
Eq.~(\ref{bose1}) in the presence of a thermal bath, or the bosonic
Boltzmann operator when elastic collisions are taken into account. An
even more general model could take into account both contributions, or
include additional semiclassical contributions arising from the fact
that the operators ${\bf r}$ and ${\bf k}$ do not commute.

As a final comment, we may note that there exists analogies between
the Bose-Einstein condensation and the inverse cascade process in
two-dimensional turbulence where energy accumulates at large scales
($k=0$) to form a macrovortex \cite{kraichnan}.

\vskip1cm
{\it Acknowledgements} This study was initiated by one of us (P.H.C)
during the 3rd NEXT-Sigma-Phi Conference in Crete (August 2005). He
thanks C. Tsallis, G. Kaniadakis and J. Sommeria for stimulating
discussions.

\end{document}